\documentclass[a4paper, UKenglish, colorlinks]{lipics-v2021}

\nolinenumbers 

\usepackage{amsfonts}
\usepackage{amsmath}
\usepackage{amsthm}
\usepackage{cite} 
\usepackage[T1]{fontenc}
\usepackage{mdwlist}
\usepackage{prooftree}
\usepackage{sourcecodepro}

\hypersetup{colorlinks,linkcolor={blue},citecolor={blue},urlcolor={red}}

\newcommand{\deffont}[1]{\textbf{#1}}
\newcommand{\fall}[1]{\forall{#1}.\ }
\newcommand{\lam}[1]{\lambda{#1}.\ }
\newcommand{\rulefont}[1]{\ensuremath{(\mathbf{#1})}}
\newcommand{\xsts}[1]{\exists{#1}.\ }

\newcommand{\krulevariable}{\rulefont{Kv}}
\newcommand{\kruleformer}{\rulefont{Kf}}
\newcommand{\kruleapplication}{\rulefont{Ka}}

\newcommand{\trulevariable}{\rulefont{Tv}}
\newcommand{\truleconstant}{\rulefont{Tc}}
\newcommand{\truleapplication}{\rulefont{Ta}}
\newcommand{\trulelambda}{\rulefont{Tl}}

\newcommand{\lrulerefl}{\rulefont{Lrefl}}
\newcommand{\lrulesym}{\rulefont{Lsym}}
\newcommand{\lruletrans}{\rulefont{Ltrns}}
\newcommand{\lruleidem}{\rulefont{Lidm}}
\newcommand{\lruleassoc}{\rulefont{Lassc}}
\newcommand{\lrulecong}{\rulefont{Lcng}}
\newcommand{\lrulebase}{\rulefont{Lbase}}

\newcommand{\nruleassm}{\rulefont{Ninit}}
\newcommand{\nruletrueintro}{\rulefont{NtrueI}}
\newcommand{\nrulefalseelim}{\rulefont{NfalseE}}
\newcommand{\nrulelift}{\rulefont{Nlift}}
\newcommand{\nrulerefl}{\rulefont{Nrefl}}
\newcommand{\nrulesym}{\rulefont{Nsym}}
\newcommand{\nruletrans}{\rulefont{Ntrans}}
\newcommand{\nrulelamcong}{\rulefont{Nlcong}}
\newcommand{\nruleappcong}{\rulefont{Nacong}}
\newcommand{\nrulebeta}{\rulefont{Nbeta}}
\newcommand{\nruletysubst}{\rulefont{Ninst}}
\newcommand{\nruleeta}{\rulefont{Neta}}
\newcommand{\nrulesubst}{\rulefont{Nsubst}}
\newcommand{\nruleiffEL}{\rulefont{NiffE1}}
\newcommand{\nruleiffER}{\rulefont{NiffE2}}
\newcommand{\nruleiffI}{\rulefont{NiffI}}
\newcommand{\nruleconjI}{\rulefont{NconjI}}
\newcommand{\nruleconjEL}{\rulefont{NconjE1}}
\newcommand{\nruleconjER}{\rulefont{NconjE2}}
\newcommand{\nruledisjIL}{\rulefont{NdisjI1}}
\newcommand{\nruledisjIR}{\rulefont{NdisjI2}}
\newcommand{\nruledisjE}{\rulefont{NdisjE}}
\newcommand{\nruleimpI}{\rulefont{NimpI}}
\newcommand{\nruleimpE}{\rulefont{NimpE}}
\newcommand{\nrulenegI}{\rulefont{NnegI}}
\newcommand{\nrulenegE}{\rulefont{NnegE}}
\newcommand{\nruleallI}{\rulefont{NallI}}
\newcommand{\nruleallE}{\rulefont{NallE}}
\newcommand{\nruleexI}{\rulefont{NexI}}
\newcommand{\nruleexE}{\rulefont{NexE}}
\newcommand{\nrulelem}{\rulefont{Nlem}}
\newcommand{\nruleraa}{\rulefont{Nraa}}
\newcommand{\nruleweak}{\rulefont{Nwk}}

\usepackage{ifthen}
\usepackage{amsmath,amssymb}
\newboolean{showcomments}
\setboolean{showcomments}{true}
\ifthenelse{\boolean{showcomments}}
  {\newcommand{\mynote}[2]{
    \fbox{\bfseries\sffamily\scriptsize#1}
    {\small$\blacktriangleright$\textsf{\emph{#2}}$\blacktriangleleft$}
   }
  }
  {\newcommand{\mynote}[2]{}
  }

\acknowledgements{The authors wish to thank Jim Grundy, John Harrison, Konrad Slind, and Christian Urban for their constructive comments on earlier drafts of this paper.}

\title{A modest proposal: explicit support for foundational pluralism}

\author{Martin Berger}{
  Montanarius Ltd, London, United Kingdom \and
  Turing Core, Huawei 2012 Labs, Huawei R\&D Ltd, London, United Kingdom \and
  University of Sussex, Brighton, United Kingdom \and \url{https://martinfriedrichberger.net}}{contact@martinfriedrichberger.net}{https://orcid.org/0000-0003-3239-5812}{}

\author{Dominic P.~Mulligan}{Automated Reasoning Group, Amazon Web Services, Cambridge, United Kingdom\footnote{All work done whilst employed by Arm Research} \and \url{https://dominicpm.github.io}}{dominic.p.mulligan@gmail.com}{https://orcid.org/0000-0003-4643-3541}{}


\Copyright{Martin Berger and Dominic P.~Mulligan}

\authorrunning{M.~Berger and D.~P.~Mulligan}


\begin{document}

\maketitle

\begin{abstract}
Whilst mathematicians assume classical reasoning principles by default they often \emph{context switch} when working, restricting themselves to various forms of subclassical reasoning.
This pattern is especially common amongst logicians and set theorists, but workaday mathematicians also commonly do this too, witnessed by narrative notes accompanying a proof---``the following proof is constructive'', or ``the following proof does not use choice'', for example.
Yet, current proof assistants provide poor support for capturing these narrative notes formally, an observation that is especially true of systems based on Gordon's HOL, a classical higher-order logic.
Consequently, HOL and its many implementations seem ironically more committed to classical reasoning than mainstream mathematicians are themselves, limiting the mathematical content that one may easily formalise.

To facilitate these context switches, we propose that mathematicians mentally employ a simple tainting system when temporarily working subclassically---an idea not currently explored in proof assistants.
As such, we introduce a series of modest but far-reaching changes to HOL, extending the standard two-place Natural Deduction relation to incorporate a \emph{taint-label}, taken from a particular lattice, and which describes or limits the ``amount'' of classical reasoning used within a proof.
Taint can be seen either as a simple typing system on HOL proofs, or as a form of static analysis on proof trees, and \emph{partitions} our logic into various fragments of differing expressivity, sitting side-by-side.
Results may pass from a ``less classical'' fragment into a ``more classical'' fragment of the logic without modification, but not \emph{vice versa}, with the flow of results between worlds controlled by an inference rule akin to a subtyping or subsumption rule.
Denizens of all worlds reason over the same set of definitions, and to maximise reuse users are therefore given incentive to phrase results in the weakest possible world that will suffice, eschewing needlessly-classical reasoning.
\end{abstract}

\section{Introduction}
\label{sect.introduction}

Mathematicians work with classical reasoning principles by default though will freely ``context switch'' between different \emph{foundational}, often \emph{subclassical}, styles of reasoning.

Logicians, set theorists, and others working on foundational issues, are the most obvious representatives of this pattern, and for example may eschew working with the Axiom of Choice in favour of more restricted choice principles---like the Axioms of Dependent and Countable Choice---to investigate their consequences~\cite{jech-axiom-of-choice}.
Within constructive mathematics, there are also several competing visions of what ``constructivism'' means, and therefore debate over which reasoning principles are admissible---the Russian School of Constructivism, rooted in recursion theory is quite distinct from Intuitionism, for example~\cite{sep-mathematics-constructive}.
This pluralism is a reflection of the fact that there is no single, universally agreed upon foundational system for mathematics, but many different plausible foundations, each of independent interest.

This context switching is also seen in more mainstream, non-foundational mathematics, too.
Proofs are often accompanied by supplementary related claims in the natural language narrative: ``the following proof is constructive'', or ``the following proof does not make use of choice'', for example.\footnote{See e.g., the note accompanying Theorem 7.2 here~\cite{cuadra-existence-2022}, or the note preceding Lemma 43 here~\cite{karagila-zornian-2020}.}
These comments invite the reader to perform a temporary context switch, away from a default classical mode of reasoning, into some restricted subclassical mode for the duration of a proof.
As readers, we are essentially invited by the author to temporarily change our assumed foundational system.

These supplementary claims are interesting in their own right, as they have an ambiguous mathematical status.
If a proof is claimed to be constructive, but is not so, is the proof incorrect?
Likewise, if a proof is claimed not to rely on the Axiom of Choice, but makes appeal to Zorn's Lemma, or any of the other results known equivalent to Choice, is this proof incorrect?
In both cases, we claim proofs are indeed manifestly incorrect, and as a consequence, these claims have a kind of mathematical content, modifying theorem statements: not merely $\phi$, but $\phi$ \emph{constructively}, or $\phi$ \emph{without choice}.
Within the context of these statement-modifying claims, proofs seem to be checked by the reader using a simple mental analysis, wherein occurrences of classical reasoning principles appearing within the proof, or indirectly in lemmas, indelibly \emph{taint} the result.
When working classically in the default foundational system of mainstream mathematics this taint system is tacit, only becoming needed---and visible---when working subclassically.

Unfortunately, existing proof assistants provide poor support for capturing this foundational context switching.
This observation applies equally to proof assistants based on Type Theory and classical HOL---representing a plurality of implemented systems.

Systems based on Type Theory---Agda~\cite{DBLP:conf/tldi/Norell09}, Coq~\cite{bertot2013interactive}, or Matita~\cite{DBLP:conf/cade/AspertiRCT11}, for example---start with an Intuitionistic base logic and use various techniques to obtain a more classical foundation.
The most straightforward technique is to add classical axioms directly.
However, owing to the distinction between definitional and propositional equalities in Type Theory, classical reasoners find themselves disadvantaged as axioms without computational content potentially ``gum up'' conversion.
Other schemes for embedding classical logic include forms of G\"odel-Gentzen-style double-negation translations~\cite{goedel-dialectica, avigad-goedel-1998, Ferreira_2011} and \emph{stable embeddings}~\cite{oconnor-classical-2011}---and which clutter goals with extraneous negations or side-conditions, making them less ergonomic for use in extended work, or alternatively working within a dedicated monad\footnote{As in the Agda standard library, for example: \url{https://agda.github.io/agda-stdlib/v1.1/Relation.Nullary.Negation.html\#1727}.}, amongst many other approaches.
Whilst classical mathematics \emph{can} technically be accommodated in Intuitionistic Type Theory, practically its disciples will always be second class citizens.

Systems implementing Gordon's HOL, a classical higher-order logic---HOL4~\cite{DBLP:conf/tphol/SlindN08}, Isabelle/HOL~\cite{DBLP:conf/tphol/WenzelPN08}, or HOL Light~\cite{DBLP:conf/tphol/Harrison09a}, for example---also have problems.
In particular, HOL systems provide no easy way of restricting reasoning into some subclassical fragment of the logic.
Use of Choice, captured through a Hilbert-style indefinite description operator, \emph{\`{a} la} the $\epsilon$-calculus~\cite{sep-epsilon-calculus}, is endemic within the logic. Consequently, via Diaconescu's theorem~\cite{diaconescu-axiom, goodman-choice-1978} the HOL Boolean type---which acts as both the type of propositions and the two-element datatype---satisfies Excluded Middle.\footnote{In HOL Light, at least, this is how Excluded Middle is actually derived.}
HOL, and its many implementations, is ironically more wedded to classical reasoning than mathematicians are themselves.

The inability to restrict the prevailing mode of reasoning within HOL has practical consequences.
\emph{Smooth Infinitesimal Analysis} (SIA) is an approach to infinitesimal analysis that requires constructive reasoning to avoid inexorable collapse of the theory \cite{BellJL:priinfa}.
Fix a suggestively-named carrier set, $\mathbb{R}$, constants $\{0, 1\} \subseteq \mathbb{R}$, and assume field operations with the usual properties closed over $\mathbb{R}$, and further define the set, $\Delta$, of \emph{nilsquare nilpotents}, where $\Delta = \{ \epsilon \in \mathbb{R} \mid \epsilon \cdot \epsilon = 0 \}$.
Assuming Excluded Middle, $\Delta$ provably collapses into the singleton set $\{ 0 \}$.
Yet, if we avoid Excluded Middle, we also avoid this collapse, and the most that can be said is that $0 \in \Delta$ and that for any $\epsilon \in \Delta$ we have $\epsilon \cdot \epsilon = 0$, leaving open the possibility that non-zero elements of $\Delta$ exist.
Taking $\Delta$ as a set of infinitesimals---elements so small that they are zero when squared---we may assume further axioms asserting the \emph{continuity} of all functions $f : \mathbb{R} \rightarrow \mathbb{R}$.
With this, we may develop analysis synthetically using infinitesimals.
If we maintain a constructive pretence, anyway.

Yet, SIA is hard to formalise ergonomically in systems implementing HOL owing to HOL's commitment to classical mathematics.
Focussing on Isabelle/HOL---the system with which the authors are familiar---we must engage in tricks, for example working within an inner deep-embedding of Intuitionistic Logic, to work with SIA.
Such working inside a \emph{theorem prover within a theorem prover} is clearly undesirable, and a departure from how SIA is introduced in practice.
Bell's introductory overview of SIA \cite{BellJL:priinfa}, for example, merely states that Excluded Middle and other classical reasoning principles will be eschewed for the duration of the book---a narrative note applied to an entire book rather than a single proof.

To conclude, our thesis is that current proof assistants are \emph{too} assertive in their foundational convictions.
Moving between reasoning styles, even the major schools of mathematical thought, is awkward, and requires compromises by working with strange or impractical encodings and embeddings.
These dogmatic convictions are out of kilter with mathematical practice, and have also contributed, in part, toward a splintering of the formalised mathematics community into disjoint subcommunities, each associated with the foundational system to which they pledge their fealty: constructivists constellated around one family of tools, classical mathematicians around another.
We make the following contributions:
\begin{enumerate*}
\item
We observe the ``context switching'' phenomenon that is not well modelled by proof checking software, leading to difficulties in naturally formalising some mathematics.
\item
In \S\ref{sect.a.more.pluralistic.hol} we make modest but far-reaching modifications to Gordon's HOL to more accurately model these context switches.
We extend the Natural Deduction relation to also include a taint-label, capturing the amount of classical reasoning used within a proof. Initially, the logic is parametric in a generic lattice (bounded from below) of taint-labels, which we later specialise with a lattice corresponding to certain mathematical systems of interest.
\item
In \S\ref{sect.data.and.definitional.principles} we add datatypes to our logic and discuss definitional principles.
We discuss refining existing HOL definitional principles, and we speculatively describe another potential principle which moves us from a fixed set of taint-labels to a logic where these can be added dynamically, thereby introducing an ``open world'' of taint.
\item
In \S\ref{sect.mathematics.and.smooth.infinitesimal.analysis} we discuss the likely pragmatic effects that our modifications to HOL have on formalised mathematics encoded within the system.
\item
In \S\ref{sect.computer.implementation} we discuss a prototype implementation of our logic in an LCF-style proof-checking system, written in Scala.
\end{enumerate*}

\section{A more pluralistic HOL}
\label{sect.a.more.pluralistic.hol}

We present a logic where users of different foundational stripes can work side-by-side, ideally on an as-equal footing as possible.
We take HOL as our base logic, with some small but important changes.
HOL has the great advantage that it includes only a single kind of equality, with no internal notion of computation with which different axioms can interfere.
In principle, the idea of tainting proofs can be applied to any other logical system.

In \S\ref{subsect.natural.deduction} we introduce an extension of the standard dyadic Natural Deduction relation, introducing a ternary relation $\Gamma \vdash \phi : \ell$ betwixt context, formula, and taint-label, and further develop the metatheory of this relation.
However, before then, we introduce basic, standard material in \S\ref{subsect.language.kinds.types.and.terms}.
We omit proofs, which are straightforward, and should not surprise anybody who is anyway familiar with Proof or Type Theory.
The cognoscenti may freely skip ahead.

\subsection{Language: kinds, types, and terms}
\label{subsect.language.kinds.types.and.terms}

We first define \deffont{kinds} by the following recursive grammar:
\begin{displaymath}
\kappa, \kappa', \kappa'' ::= \star \mid \star \Rightarrow \kappa
\end{displaymath}
The \deffont{kind arrow} associates to the right, and we write $\star \Rightarrow \star \Rightarrow \star$ as a shorthand for $\star \Rightarrow (\star \Rightarrow \star)$.
Kinds are restricted in form, with the \deffont{kind of types}, $\star$, always appearing to the left of a kind arrow, $- {\Rightarrow} -$, making kinds isomorphic to the Natural Numbers.

We also fix a countably infinite set of \deffont{type-variables}, and use $\alpha$, $\beta$, $\gamma$, and so on, to range arbitrarily over type-variables.
For type-variables, we employ a \deffont{permutative convention} wherein type-variables with distinct names are assumed to be distinct, so that $\alpha \not= \beta$ always, a trick that will simplify some theorem statements later.

To each kind, $\kappa$, we associate a countably infinite set of \deffont{type-formers}, and we use $\mathtt{F}{:}\kappa$, $\mathtt{G}{:}\kappa$, $\mathtt{H}{:}\kappa$, and so on, to range arbitrarily over the type-formers associated with the kind $\kappa$.
Later, we may sometimes drop the kind annotation on a type-former, preferring to write $\mathtt{F}$ instead of $\mathtt{F}{:}\kappa$, though when we do the kind will be inferable from context.
If $\kappa \not= \kappa'$ then there is no connection between $\mathtt{F}{:}\kappa$ and $\mathtt{F}{:}\kappa'$, though we will avoid this sort of name clash.

Henceforth, associated with the kind $\star$, we assume that the infinite set of type-formers contains \emph{at least} the distinguished type-former $\mathtt{Prop}$, and similarly assume that the infinite set of type-formers associated with the kind $\star \Rightarrow \star \Rightarrow \star$ contains \emph{at least} the distinguished type-former $- {\rightarrow} -$.
More type-formers will be assumed later.
Here, we note an early distinction between our logic and HOL, as we cleave the HOL Boolean type into two types, each dedicated to a different purpose.
We use $\mathtt{Prop}$ as our type of propositions.
Later, we introduce a Boolean type, $\mathtt{Bool}$, as a two-element datatype, with usual recursion and induction principle, like any other.
HOL, by treating the type of propositions as data, introduces yet another backdoor through which Excluded Middle may enter, as the principle is immediately derivable from the Boolean type's structural induction rule.

Next, we define \deffont{pre-types} by the following recursive grammar:
\begin{displaymath}
\tau, \tau', \tau'' ::= \alpha \mid \mathtt{F}{:}\kappa \mid \tau\tau'
\end{displaymath}
We write $ftv(\tau)$ for the set of \deffont{free type-variables} appearing in the (pre-)type $\tau$, and write $\tau[\beta := \tau']$ for the \deffont{type-substitution} of pre-type $\tau'$ for $\beta$ in $\tau$, satisfying:

\begin{lemma}
\label{lemm.pre-type.type-variable.substitution.identity}
$\tau[\alpha := \alpha] = \tau$
\end{lemma}

\begin{lemma}
\label{lemm.pre-type.type-variable.substitution.garbage.collection}
If $\beta \notin ftv(\tau)$ then $\tau[\beta := \tau'] = \tau$.
\end{lemma}

\begin{lemma}
\label{lemm.pre-type.type-variable.substitution.commute}
If $\beta \notin ftv(\tau'')$ then $\tau[\beta := \tau'][\gamma := \tau''] = \tau[\gamma := \tau''][\beta := \tau'[\gamma := \tau'']]$.
\end{lemma}

\begin{lemma}
\label{lemm.pre-type.type-variable.substitution.free.type-variable.reducing}
$ftv(\tau[\beta := \tau']) \subseteq (ftv(\tau) - \{\beta\}) \cup ftv(\tau')$
\end{lemma}

We define a \deffont{kinding relation} on pre-types using the following rules, and write $\vdash \tau : \kappa$ to assert that a derivation tree exists, constructed per the rules below, and rooted at $\vdash \tau : \kappa$:
\begin{gather*}
\begin{prooftree}
\phantom{h}
\justifies
\vdash \alpha : \star
\using\krulevariable
\end{prooftree}
\qquad
\begin{prooftree}
\phantom{h}
\justifies
\vdash \mathtt{F}{:}\kappa : \kappa
\using\kruleformer
\end{prooftree}
\qquad
\begin{prooftree}
\vdash \tau : \star \Rightarrow \kappa
\quad
\vdash \tau' : \star
\justifies
\vdash \tau\tau' : \kappa
\using\kruleapplication
\end{prooftree}
\end{gather*}
If $\vdash \tau : \star$ then we call $\tau$ a \deffont{type}, and in the remainder of the paper, we will generally work exclusively with types.
Kinding enjoys a few ``obvious'' correctness properties:
\begin{lemma}
\label{lemm.pre-type.type-variable.substitution.preserves.kinding}
If $\vdash \tau : \kappa$ and $\vdash \tau' : \star$ then $\vdash \tau[\beta := \tau'] : \kappa$.
\end{lemma}

\begin{lemma}
\label{lemm.kinding.relation.unicity}
If $\vdash \tau : \kappa$ and $\vdash \tau : \kappa'$ then $\kappa = \kappa'$.
\end{lemma}
We abuse syntax somewhat, and write $\mathtt{Prop}$ for the \deffont{type of propositions}, as well as the underlying type-former.
Moreover, whenever $\tau$ and $\tau'$ are types then $\tau \rightarrow \tau'$ is a type, also---we call this a \deffont{function type}.
The function arrow associates to the right, and we may write $\tau \rightarrow \tau' \rightarrow \tau''$ as a shorthand for $\tau \rightarrow (\tau' \rightarrow \tau'')$.

Now, to each pre-type, $\tau$, we associate a countably infinite set of \deffont{term variables}---or just \deffont{variables} when no confusion between type and term variables is likely---and a countably infinite set of \deffont{constants}.
We use $x{:}\tau$, $y{:}\tau$, $z{:}\tau$, and so on, and $\mathtt{C}{:}\tau$, $\mathtt{D}{:}\tau$, $\mathtt{E}{:}\tau$, and so on, to range arbitrarily over variables and constants, respectively, associated with pre-type $\tau$.
Again we may drop type annotations when convenient, though the pre-type $\tau$ will always be inferable whenever we do this.
Moreover, if $\tau \not= \tau'$ then there is no particular connection between the variables $x{:}\tau$ and $x{:}\tau'$ nor constants $C{:}\tau$ and $C{:}\tau'$.
For variables, we again employ a \deffont{permutative convention}, and we assume the following constants, corresponding to the logical connectives and quantifiers, with associated types:
\begin{displaymath}
\begin{tabular}{ccl}
$\top$, $\bot$ &  & $\mathtt{Prop}$ \\
$\wedge$, $\vee$, $\longrightarrow$, $\longleftrightarrow$ & & $\mathtt{Prop} \rightarrow \mathtt{Prop} \rightarrow \mathtt{Prop}$ \\
$\neg$ & \text{ with type } & $\mathtt{Prop} \rightarrow \mathtt{Prop}$ \\
$=$ & & $\alpha \rightarrow \alpha \rightarrow \mathtt{Prop}$ \\
$\exists$, $\forall$ & & $(\alpha \rightarrow \mathtt{Prop}) \rightarrow \mathtt{Prop}$
\end{tabular}
\end{displaymath}
The eagle-eyed will observe another difference between our logic and HOL: the lack of any analogue of the Hilbert $\epsilon$-operator.
This is a mechanism by which one may \emph{choose} an element satisfying a predicate, with the term $\epsilon x. P\ x$ either denoting an element satisfying $P$, or remaining undefined if no such element exists.
To imbue this construct with meaning, HOL implementations typically use an additional Natural Deduction rule:
\begin{displaymath}
\begin{prooftree}
\Gamma \vdash \xsts{x}P\ x
\justifies
\Gamma \vdash P\ (\epsilon x. P\ x) 
\end{prooftree}
\end{displaymath}
Intuitively: the property $P$ holds of the element selected by the $\epsilon$-operator if we can first show that such an element exists.
Using this, one can \emph{define} the inverse of a function, and reason about it, and also \emph{prove} the equivalent HOL statement of the Axiom of Choice.

For our purposes, this is problematic.
Hilbert's $\epsilon$-operator is, in a sense, \emph{too} expressive as one may use it to derive the Axiom of Choice, as just observed, and would preclude restricting reasoning to any of the zoo of weaker choice principles.
Further, if we include the $\epsilon$-operator as a constant, we must decide what to do with definitions using this construct when we are restricted to subclassical reasoning.
One approach would be to leave the $\epsilon$-operator as uninterpreted, restricting usage of the rule presented above, when working subclassically.
From an aesthetic perspective, this seems rather ugly.
We therefore dispense with description operators and instead introduce choice as an explicit \emph{axiom}.
This decision has a marked effect on how one uses our logic, a subject which we will return to, later in the paper, in \S\ref{sect.mathematics.and.smooth.infinitesimal.analysis}.

We recursively define explicitly-typed \deffont{$\lambda$-terms} via the following grammar:
\begin{gather*}
r,s,t ::= x{:}\tau \mid \mathtt{C}{:}\tau \mid rs \mid \lam{x{:}\tau}r
\end{gather*}
As usual, the variable $x{:}\tau$ is said to be \deffont{bound} in the term $\lam{x{:}\tau}r$, and we henceforth work with terms that are identified up-to $\alpha$-equivalence, with all definitions from this point onward well-defined with respect to $\alpha$-equivalence.
We take pity on the reader, and adopt common mathematical conventions, writing, for example, $\fall{x{:}\tau}\phi$ instead of $\forall[\alpha := \tau](\lam{x{:}\tau}\phi)$, and $\phi \longrightarrow \psi$ instead of $(\longrightarrow\phi)\psi$.
Similarly for the other logical connectives and quantifiers.

We write $fv(r)$ for the set of \deffont{free variables} appearing in $r$, so that $fv(\lam{x{:}\tau}y{:}\tau') = \{ y{:}\tau' \}$, and write $ftv(r)$ for the \deffont{type variables} appearing in $r$.
We write $r[\alpha := \tau]$ for the extension of type substitution to a \deffont{type substitution action} on terms, and write $r[x{:}\tau := s]$ for the \deffont{capture-avoiding substitution action} which replaces all occurrences of $x{:}\tau$ by $s$ in $r$ whilst renaming bound variables as appropriate.
The type substitution action satisfies some obvious properties:

\begin{lemma}
\label{lemm.term.type-variable.substitution.identity}
$r[\beta := \beta] = r$
\end{lemma}

\begin{lemma}
\label{lemm.term.type-variable.substitution.garbage.collection}
If $\beta \notin ftv(r)$ then $r[\beta := \tau'] = r$.
\end{lemma}

\begin{lemma}
\label{lemm.term.type-variable.substitution.commute}
If $\beta \notin ftv(\tau'')$ then $r[\beta := \tau'][\gamma := \tau''] = r[\gamma := \tau''][\beta := \tau'[\gamma := \tau'']]$.
\end{lemma}

\begin{lemma}
\label{lemm.term.type-variable.substitution.free.type-variable.reducing}
$ftv(r[\beta := \tau']) \subseteq (ftv(r) - \{ \beta \}) \cup ftv(\tau')$
\end{lemma}

\noindent
Likewise, the capture-avoiding substitution action:

\begin{lemma}
\label{lemm.term.capture-avoiding.substitution.identity}
$r[x{:}\tau := x{:}\tau] = r$
\end{lemma}

\begin{lemma}
\label{lemm.term.capture-avoiding.substitution.garbage.collect}
If $x{:}\tau \notin fv(r)$ then $r[x{:}\tau := s] = r$.
\end{lemma}

\begin{lemma}
\label{lemm.term.capture-avoiding.substitution.reduces.free.variables}
$fv(r[x{:}\tau := s]) \subseteq (fv(r) - \{ x{:}\tau \}) \cup fv(s)$
\end{lemma}

\begin{lemma}
\label{lemm.term.capture-avoiding.substitution.commute}
If $x{:}\tau \notin fv(t)$ then $r[x{:}\tau := s][y{:}\tau' := t] = r[y{:}\tau' := t][x{:}\tau := s[y{:}\tau' := t]]$.
\end{lemma}

We introduce a \deffont{typing relation} on terms, defined by the rules below, and write $\vdash r : \tau$ to assert that a derivation tree exists, constructed per the rules below, and rooted at $\vdash r : \tau$:
\begin{gather*}
\begin{prooftree}
\vdash \tau : \star
\justifies
\vdash x{:}\tau : \tau
\using\trulevariable
\end{prooftree}
\quad
\begin{prooftree}
\vdash \tau : \star
\justifies
\vdash \mathtt{C}{:}\tau : \tau
\using\truleconstant
\end{prooftree}
\quad
\begin{prooftree}
\vdash r : \tau \rightarrow \tau'
\quad
\vdash s : \tau
\justifies
\vdash rs : \tau'
\using\truleapplication
\end{prooftree}
\quad
\begin{prooftree}
\vdash r : \tau'
\quad
\vdash \tau : \star
\justifies
\vdash \lam{x{:}\tau}r : \tau \rightarrow \tau'
\using\trulelambda
\end{prooftree}
\end{gather*}
If $\vdash r : \tau$ for some $\tau$ then we call $r$ \deffont{well-typed}---again, generally speaking, we will always work with well-typed terms.
Moreover, if $\vdash r : \mathtt{Prop}$ then we call $r$ a \deffont{formula}, using $\phi$, $\psi$, $\xi$, and so on, to range arbitrarily over terms that we wish to suggest should be understood as formulae.
Again, this typing relation satisfies some obvious correctness properties:

\begin{lemma}
\label{lemm.typing.relation.implies.kinding}
If $\vdash r : \tau$ then $\vdash \tau : \star$.
\end{lemma}

\begin{lemma}
\label{lemm.typing.relation.unicity}
If $\vdash r : \tau$ and $\vdash r : \tau'$ then $\tau = \tau'$.
\end{lemma}

\begin{lemma}
\label{lemm.typing.relation.type-variable.substitution}
If $\vdash r : \tau$ and $\vdash \tau' : \star$ then $\vdash r[\beta := \tau'] : \tau[\beta := \tau']$.
\end{lemma}

\begin{lemma}
\label{lemm.typing.relation.capture-avoiding.substitution}
If $\vdash r : \tau$ and $\vdash s : \tau'$ then $\vdash r[y{:}\tau' := s] : \tau$.
\end{lemma}

Lastly, note that equality at type $\mathtt{Prop}$ coincides with bi-implication.
We assume an additional constant $\longleftrightarrow$ of type $\mathtt{Prop} \rightarrow \mathtt{Prop} \rightarrow \mathtt{Prop}$, and write $\phi \longleftrightarrow \psi$ instead of $\phi = \psi$.

\subsection{Natural Deduction}
\label{subsect.natural.deduction}

Our most notable deviation from HOL is the introduction of taint tracking, formalising the informal process described in \S\ref{sect.introduction}.
HOL is typically presented in Natural Deduction form, with the Natural Deduction relation a dyadic relation between contexts and formulae, $\Gamma \vdash \phi$.
We extend this relation to a ternary relation, between context, formula, and \emph{taint-label}, $\Gamma \vdash \phi : \ell$.

We will introduce our extension of Natural Deduction by working with a core logic constructed over an arbitrary lattice.
The choice of which \emph{foundationally interesting axioms}\footnote{We remain ambivalent about what a \emph{foundationally interesting axiom} actually is, but suggest they should vary the universe of mathematical objects, or what can be deduced about them, in some interesting way.} used to ``partition'' the logic is a matter of taste, and reflects the mathematical reasoning styles of interest to users.
Axioms that introduce taint can be varied as long as the axioms and the lattice that they generate satisfy certain properties.
Despite this, we will later extend the logic with axioms capturing a few interesting systems of mathematics, and also discuss a new definitional principle for dynamically adding axioms and associated taint-labels.

To do this, we first fix a set of \deffont{taint-labels}, $\mathcal{L}$, and use $\ell$, $\ell'$, $\ell''$, and so on, to range arbitrarily over taint-labels.
We also fix a \deffont{binary operation}, $- \sqcup -$, on taint-labels with a \deffont{closure} property, so that $\ell \sqcup \ell' \in \mathcal{L}$ whenever $\ell \in \mathcal{L}$ and $\ell' \in \mathcal{L}$.
We assume that the set of taint-labels, $\mathcal{L}$, contains \emph{at least} the distinguished label $I$, though more taint-labels will be assumed later.
We also introduce a notion of \deffont{derivable equivalence} between taint-labels, with the rules in Figure~\ref{fig.taint-label.equational.theory}.
We write $\vdash \ell \equiv \ell'$ to assert that a derivation tree exists, constructed per the rules in Figure~\ref{fig.taint-label.equational.theory}, and rooted at $\vdash \ell \equiv \ell'$.

\begin{figure}[ht]
\begin{gather*}
\begin{prooftree}
\phantom{h}
\justifies
\vdash \ell \equiv \ell
\using\lrulerefl
\end{prooftree}
\quad
\begin{prooftree}
\vdash \ell \equiv \ell'
\justifies
\vdash \ell' \equiv \ell
\using\lrulesym
\end{prooftree}
\quad
\begin{prooftree}
\vdash \ell \equiv \ell'
\quad
\vdash \ell' \equiv \ell''
\justifies
\vdash \ell \equiv \ell''
\using\lruletrans
\end{prooftree}
\quad
\begin{prooftree}
\phantom{h}
\justifies
\vdash \ell \sqcup \ell \equiv \ell
\using\lruleidem
\end{prooftree}
\\[1.5ex]
\begin{prooftree}
\phantom{h}
\justifies
\vdash \ell \sqcup (\ell' \sqcup \ell'') \equiv (\ell \sqcup \ell') \sqcup \ell''
\using\lruleassoc
\end{prooftree}
\quad
\begin{prooftree}
\vdash \ell \equiv \ell'
\quad
\vdash \ell'' \equiv \ell'''
\justifies
\vdash \ell \sqcup \ell'' \equiv \ell' \sqcup \ell'''
\using\lrulecong
\end{prooftree}
\quad
\begin{prooftree}
\phantom{h}
\justifies
\vdash I \sqcup \ell \equiv \ell
\using\lrulebase
\end{prooftree}
\end{gather*}
\caption{The equational theory between taint-labels}
\label{fig.taint-label.equational.theory}
\end{figure}

The rules above imbue taint-labels with a \deffont{lattice structure}, presented here in an algebraic form.
Like all lattices, we can also present this in an order-theoretic manner, and we write $\vdash \ell \leq \ell'$ to assert that $\vdash \ell \sqcup \ell' \equiv \ell'$.
We call this relation the \deffont{derived ordering} on taint-labels.
Both algebraic and order-theoretic presentations will be useful.
This lattice is \deffont{bounded from below}, with $I$ the least element with respect to this derived order, in the sense that $\vdash I \leq \ell$ for all taint-labels, $\ell$.
The label $I$ represents purely-constructive reasoning.
This order satisfies various properties:

\begin{lemma}
\label{lemm.derived.ordering.subsumes.equivalence}
If $\vdash \ell \equiv \ell'$ then $\vdash \ell \leq \ell'$.
\end{lemma}

\begin{lemma}
\label{lemm.derived.ordering.transitive}
If $\vdash \ell \leq \ell'$ and $\vdash \ell' \leq \ell''$ then $\vdash \ell \leq \ell''$.
\end{lemma}

\begin{lemma}
\label{lemm.derived.ordering.antisymmetric}
If $\vdash \ell \leq \ell'$ and $\vdash \ell' \leq \ell$ then $\vdash \ell \equiv \ell'$.
\end{lemma}

\begin{lemma}
\label{lemm.derived.ordering.respects.taint.label.binary.operation}
If $\vdash \ell \leq \ell'$ and $\vdash \ell'' \leq \ell'''$ then $\vdash \ell \sqcup \ell'' \leq \ell' \sqcup \ell'''$.
\end{lemma}

We call a finite set of terms a \deffont{context}, and use $\Gamma$, $\Gamma'$, $\Gamma''$, and so on, to range arbitrarily over contexts.
We call a context, $\Gamma$, \deffont{valid} whenever $\vdash \phi : \mathtt{Prop}$ for every $\phi \in \Gamma$, and write $\vdash \Gamma \text{ valid}$ to assert that the context $\Gamma$ is valid.

\begin{lemma}
\label{lemm.context.validity.basic.properties}
We have:
\begin{enumerate}
\item
$\vdash \{\} \text{ valid}$,
\item
If $\vdash \Gamma \text{ valid}$ and $\vdash \Gamma' \text{ valid}$ then $\vdash \Gamma \cup \Gamma' \text{ valid}$.
\end{enumerate}
\end{lemma}

We write $fv(\Gamma)$ for the \deffont{free-variables} of the context, $\Gamma$, namely the set $\bigcup\{ fv(\phi) \mid \phi \in \Gamma \}$.
We also define a \deffont{pointwise type-substitution action} on contexts, $\Gamma[\beta := \tau]$ by $\Gamma[\beta := \tau] = \{ \phi[\beta := \tau] \mid \phi \in \Gamma \}$.

\begin{lemma}
\label{lemm.context.validity.preserved.by.term.type-variable.substitution}
If $\vdash \Gamma \text{ valid}$ and $\vdash \tau : \star$ then $\vdash \Gamma[\beta := \tau] \text{ valid}$.
\end{lemma}

\begin{figure*}
\begin{gather*}
\begin{prooftree}
\vdash \Gamma \text{ valid}
\quad
\phi \in \Gamma
\justifies
\Gamma \vdash \phi : I
\using\nruleassm
\end{prooftree}
\quad
\begin{prooftree}
\vdash \Gamma \text{ valid}
\justifies
\Gamma \vdash \top : I
\using\nruletrueintro
\end{prooftree}
\quad
\begin{prooftree}
\Gamma \vdash \bot : \ell
\quad
\vdash \phi : \mathtt{Prop}
\justifies
\Gamma \vdash \phi : \ell
\using\nrulefalseelim
\end{prooftree}
\\[1.5ex]
\begin{prooftree}
\Gamma \vdash \phi : \ell
\quad
\vdash \ell \leq \ell'
\justifies
\Gamma \vdash \phi : \ell'
\using\nrulelift
\end{prooftree}
\quad
\begin{prooftree}
\vdash \Gamma \text{ valid}
\quad
\vdash r : \tau
\justifies
\Gamma \vdash r = r : I
\using\nrulerefl
\end{prooftree}
\quad
\begin{prooftree}
\Gamma \vdash r = s : \ell
\justifies
\Gamma \vdash s = r : \ell
\using\nrulesym
\end{prooftree}
\\[1.5ex]
\begin{prooftree}
\Gamma \vdash r = s : \ell
\quad
\Gamma \vdash s = t : \ell
\justifies
\Gamma \vdash r = t : \ell
\using\nruletrans
\end{prooftree}
\quad
\begin{prooftree}
\Gamma \vdash r = s : \ell
\quad
x{:}\tau \notin fv(\Gamma)
\justifies
\Gamma \vdash \lam{x{:}\tau}r = \lam{x{:}\tau}s : \ell
\using\nrulelamcong
\end{prooftree}
\\[1.5ex]
\begin{prooftree}
\Gamma \vdash f = g : \ell
\quad
\Gamma \vdash r = s : \ell
\quad
\vdash f r : \tau'
\justifies
\Gamma \vdash fr = gs : \ell
\using\nruleappcong
\end{prooftree}
\quad
\begin{prooftree}
\Gamma \vdash \phi : \ell
\quad
\vdash r : \tau
\justifies
\Gamma[x{:}\tau := r] \vdash \phi[x{:}\tau := r] : \ell
\using\nrulesubst
\end{prooftree}
\\[1.5ex]
\begin{prooftree}
\vdash \Gamma \text{ valid}
\quad
\vdash \lam{x{:}\tau}r : \tau \rightarrow \tau'
\quad
\vdash s : \tau
\justifies
\Gamma \vdash (\lam{x{:}\tau}r)s = r[x{:}\tau := s] : I
\using\nrulebeta
\end{prooftree}
\quad
\begin{prooftree}
\Gamma \vdash \phi : \ell
\quad
\vdash \tau : \star
\justifies
\Gamma[\beta := \tau] \vdash \phi[\beta := \tau] : \ell
\using\nruletysubst
\end{prooftree}
\\[1.5ex]
\begin{prooftree}
\vdash \Gamma \text{ valid}
\quad
\vdash \lam{x{:}\tau}(fx) : \tau \rightarrow \tau'
\quad
x{:}\tau \notin fv(f)
\justifies
\Gamma \vdash \lam{x{:}\tau}(fx) = f : I
\using\nruleeta
\end{prooftree}
\quad
\begin{prooftree}
\Gamma \cup \{ \phi \} \vdash \bot : \ell
\justifies
\Gamma \vdash \neg\phi : \ell
\using\nrulenegI
\end{prooftree}
\\[1.5ex]
\begin{prooftree}
\Gamma \vdash \phi : \ell
\quad
\Gamma \vdash \neg\phi : \ell
\justifies
\Gamma \vdash \bot : \ell
\using\nrulenegE
\end{prooftree}
\quad
\begin{prooftree}
\Gamma \vdash \phi \longleftrightarrow \psi : \ell
\quad
\Gamma \vdash \phi : \ell
\justifies
\Gamma \vdash \psi : \ell
\using\nruleiffEL
\end{prooftree}
\quad
\begin{prooftree}
\Gamma \vdash \phi : \ell
\quad
\vdash \psi : \mathtt{Prop}
\justifies
\Gamma \cup \{ \psi \} \vdash \phi : \ell
\using\nruleweak
\end{prooftree}
\\[1.5ex]
\begin{prooftree}
\Gamma \vdash \phi \longleftrightarrow \psi : \ell
\quad
\Gamma \vdash \psi : \ell
\justifies
\Gamma \vdash \phi : \ell
\using\nruleiffER
\end{prooftree}
\quad
\begin{prooftree}
\Gamma \cup \{ \psi \} \vdash \phi : \ell
\quad
\Gamma \cup \{ \phi \} \vdash \psi : \ell
\justifies
\Gamma \vdash \phi \longleftrightarrow \psi : \ell
\using\nruleiffI
\end{prooftree}
\\[1.5ex]
\begin{prooftree}
\Gamma \vdash \phi : \ell
\quad
\Gamma \vdash \psi : \ell
\justifies
\Gamma \vdash \phi \wedge \psi : \ell
\using\nruleconjI
\end{prooftree}
\quad
\begin{prooftree}
\Gamma \vdash \phi \wedge \psi : \ell
\justifies
\Gamma \vdash \phi : \ell
\using\nruleconjEL
\end{prooftree}
\quad
\begin{prooftree}
\Gamma \vdash \phi \wedge \psi : \ell
\justifies
\Gamma \vdash \psi : \ell
\using\nruleconjER
\end{prooftree}
\\[1.5ex]
\begin{prooftree}
\Gamma \vdash \phi : \ell
\quad
\vdash \psi : \mathtt{Prop}
\justifies
\Gamma \vdash \phi \vee \psi : \ell
\using\nruledisjIL
\end{prooftree}
\quad
\begin{prooftree}
\Gamma \vdash \phi \vee \psi : \ell
\quad
\Gamma \cup \{ \phi \} \vdash \xi : \ell
\quad
\Gamma \cup \{ \psi \} \vdash \xi : \ell
\justifies
\Gamma \vdash \xi : \ell
\using\nruledisjE
\end{prooftree}
\\[1.5ex]
\begin{prooftree}
\Gamma \vdash \psi : \ell
\quad
\vdash \phi : \mathtt{Prop}
\justifies
\Gamma \vdash \phi \vee \psi : \ell
\using\nruledisjIR
\end{prooftree}
\quad
\begin{prooftree}
\Gamma \cup \{ \phi \} \vdash \psi : \ell
\justifies
\Gamma \vdash \phi \longrightarrow \psi : \ell
\using\nruleimpI
\end{prooftree}
\quad
\begin{prooftree}
\Gamma \vdash \phi \longrightarrow \psi : \ell
\quad
\Gamma \vdash \phi : \ell
\justifies
\Gamma \vdash \psi : \ell
\using\nruleimpE
\end{prooftree}
\\[1.5ex]
\begin{prooftree}
\Gamma \vdash \fall{x{:}\tau}\phi : \ell
\quad
\vdash r : \tau
\justifies
\Gamma \vdash \phi[x{:}\tau := r] : \ell
\using\nruleallE
\end{prooftree}
\quad
\begin{prooftree}
\Gamma \vdash \phi : \ell
\quad
x{:}\tau \notin fv(\Gamma)
\justifies
\Gamma \vdash \fall{x{:}\tau}\phi : \ell
\using\nruleallI
\end{prooftree}
\quad
\begin{prooftree}
\Gamma \vdash \phi[x{:}\tau := r] : \ell
\justifies
\Gamma \vdash \xsts{x{:}\tau}\phi : \ell
\using\nruleexI
\end{prooftree}
\\[1.5ex]
\begin{prooftree}
\Gamma \vdash \xsts{x{:}\tau}\phi : \ell
\quad
\Gamma \cup \{ \phi[x{:}\tau := y{:}\tau] \} \vdash \psi : \ell
\quad
y{:}\tau \notin fv(\Gamma) \cup fv(\phi)
\justifies
\Gamma \vdash \psi : \ell
\using\nruleexE
\end{prooftree}
\end{gather*}
\caption{Core rules of the Natural Deduction relation}
\label{fig.core.natural.deduction.rules}
\end{figure*}

We now define a \deffont{Natural Deduction} relation between context, term, and taint-label, $\Gamma \vdash r : \ell$, using the rules in Figure~\ref{fig.core.natural.deduction.rules}.
We write $\Gamma \vdash r : \ell$ to assert that a derivation tree exists, constructed per the rules in Figure~\ref{fig.core.natural.deduction.rules}, and rooted at $\Gamma \vdash r : \ell$.

\begin{lemma}
If $\Gamma \vdash \phi : \ell$ then $\vdash \phi : \mathtt{Prop}$.
\end{lemma}

\begin{lemma}
If $\Gamma \vdash \phi : \ell$ then $\vdash \Gamma \text{ valid}$.
\end{lemma}

\begin{lemma}
If $\Gamma \vdash \phi : \ell$ then $\ell \in \mathcal{L}$.
\end{lemma}

\begin{lemma}
If $\Gamma \vdash \phi : \ell$ and $\Gamma \subseteq \Gamma'$ and $\vdash \Gamma' \text{ valid}$ then $\Gamma' \vdash \phi : \ell$.
\end{lemma}

Intuitively, taint represents a simple static analysis over derivation trees with taint-labels ``injected'' into trees by instances of foundationally interesting axioms.
Thus far, these axioms are still missing from the rules above, but will eventually include axioms such as Excluded Middle, Choice, and similar---introducing associated taint-labels.
Inference rules of the logic are modified to allow these taint-labels to ``filter down'' through a derivation tree, emerging at the root.
The conjunction introduction rule is an instructive example:
\begin{displaymath}
\begin{prooftree}
\Gamma \vdash \phi : \ell
\quad
\Gamma \vdash \psi : \ell
\justifies
\Gamma \vdash \phi \wedge \psi : \ell
\using\nruleconjI
\end{prooftree}
\end{displaymath}
The taint-labels appearing in both premises are identical, and merely pass through the rule.
All inference rules are modified in this way, merely propagating taint-labels forward.
Any foundationally interesting axiom is therefore eventually reflected at the root of a derivation tree, justifying the name ``taint''.
To weaken taint-labels within a premise---and eventually move from a constructive proof to a classical one, for example---we also introduce an \deffont{embedding rule} strongly reminiscent of a subtyping rule:
\begin{displaymath}
\begin{prooftree}
\Gamma \vdash \phi : \ell
\quad
\vdash \ell \leq \ell'
\justifies
\Gamma \vdash \phi : \ell'
\using\nrulelift
\end{prooftree}
\end{displaymath}
Note, when designing our Natural Deduction relation, we had a design choice: rather than mandating explicit use of the embedding rule, above, to place obtain a common taint-label within the premises, we could instead phrase rules like conjunction introduction as follows:
\begin{displaymath}
\begin{prooftree}
\Gamma \vdash \phi : \ell
\quad
\Gamma \vdash \psi : \ell'
\justifies
\Gamma \vdash \phi \wedge \psi : \ell \sqcup \ell'
\end{prooftree}
\end{displaymath}
Here, $\ell \sqcup \ell'$ is the \emph{least upper bound} of the taint-labels $\ell$ and $\ell'$, itself a taint-label drawn from our lattice, and which has the effect of collecting taint together and properly propagating it through a proof.
Without extraneous appeal to the embedding rule, this system computes a \emph{precise} taint-label, describing the weakest subsystem of our logic within which a result holds, at the cost of requiring a least upper bound for every two taint-labels.\footnote{Note that this is somewhat reminiscent of Reverse Mathematics~\cite{reverse-mathematics}.}
In this system, the embedding rule is redundant: rules have weakening implicitly ``built in'', and the embedding rule \emph{commutes} with every other; derivation trees may be rewritten, in a height-decreasing fashion, so that the embedding rule only ever appears at the tree root, if at all.

However, this approach complicates tactic-driven backward proof on a computer, as a rule's premises may have a different set of taint-labels to the rule's conclusion.
In contrast, the presentation we favour has the advantage of simplifying a computer implementation, as in most cases, barring \nrulelift, taint-labels appearing in the premise of new subgoals generated by a tactic are the same as those in the original subgoal being simplified.
In this scheme, a rule's premises may always be embedded into a common fragment of the logic, owing to the existence of least-upper bounds of taint-labels.
Nevertheless, modified rules, such as the one above, and others like it, including the following alternative disjunction elimination rule
\begin{gather*}
\begin{prooftree}
\Gamma \vdash \phi \vee \psi : \ell
\quad
\Gamma \cup \{ \phi \} \vdash \xi : \ell'
\quad
\Gamma \cup \{ \psi \} \vdash \xi : \ell''
\justifies
\Gamma \vdash \xi : (\ell \sqcup \ell') \sqcup \ell'
\end{prooftree}
\end{gather*}
are all \deffont{derivable rules} within our system, owing to the embedding rule and the fact that $\ell \leq \ell \sqcup \ell'$ for all taint-labels $\ell$ and $\ell'$.
Note that the rule above is \emph{well-defined} due to the associativity of $- \sqcup -$ and the following result which demonstrates that the Natural Deduction rules respect the equational theory over taint-labels:

\begin{lemma}
If $\Gamma \vdash \phi : \ell$ and $\vdash \ell \equiv \ell'$ then $\Gamma \vdash \phi : \ell'$.
\end{lemma}

Other derived rules may also be found.
For example, the following is an analogue of the \emph{Cut} rule, useful for backwards-directed proof, and must be made ``taint aware'':

\begin{lemma}
\begin{enumerate*}
\item
If $\Gamma \vdash \phi : \ell$ and $\Gamma \cup \{ \phi \} \vdash \psi : \ell$ then $\Gamma \vdash \psi : \ell$.
\item
If $\Gamma \vdash \phi : \ell$ and $\Gamma \cup \{ \phi \} \vdash \psi : \ell'$ then $\Gamma \vdash \psi : \ell \sqcup \ell'$.
\end{enumerate*}
\end{lemma}

\subsection{Adding foundational axioms}

When conjecturing a result within our logic, a taint-label places an upper-bound on the types of reasoning that one may use over the course of a proof.
However, to observe this, we first need further taint-labels.
Assume that $\mathcal{L}$ contains an additional distinguished label, $C$, and update the equational theory of Figure~\ref{fig.taint-label.equational.theory} and the Natural Deduction relation of Figure~\ref{fig.core.natural.deduction.rules}, respectively, to introduce a new ``classical world'', by adding Excluded Middle:
\begin{gather*}
\begin{prooftree}
\vdash \Gamma \text{ valid}
\quad
\vdash \phi : \mathtt{Prop}
\justifies
\Gamma \vdash \phi \vee \neg\phi : C
\using{\nrulelem}
\end{prooftree}
\qquad
\begin{prooftree}
\phantom{h}
\justifies
\vdash C \sqcup I \equiv C
\end{prooftree}
\end{gather*}

Note that $\vdash I \leq C$ as a result of the above, which combined with the embedding rule allows results residing in the ``constructivist world'', labelled with $I$, to be lifted into the ``classical world'' labelled by $C$.
Moreover, we immediately have new reasoning principles available to denizens of the classical world, out of reach to the constructivists, for example propositional case analysis and \emph{reductio ad absurdum}, expressed as derived rules:
\begin{gather*}
\begin{prooftree}
\Gamma \cup \{ \phi \} \vdash \xi : \ell
\quad
\Gamma \cup \{ \neg\phi \} \vdash \xi : \ell
\quad
\vdash \phi : \mathtt{Prop}
\justifies
\Gamma \vdash \xi : C
\end{prooftree}
\qquad
\begin{prooftree}
\Gamma \cup \{ \neg\phi \} \vdash \bot : \ell
\justifies
\Gamma \vdash \phi : C
\using{\nruleraa}
\end{prooftree}
\end{gather*}
We introduce further foundational axioms, representing a number of commonly-axiomatised foundational systems of interest.
As above, we assume more labels, $W$ and $Ch$, and syntactically relate these to the other labels by amending the rules of Figure~\ref{fig.taint-label.equational.theory} as follows:
\begin{gather*}
\begin{prooftree}
\justifies
\vdash W \sqcup I \equiv W
\end{prooftree}
\qquad
\begin{prooftree}
\justifies
\vdash C \sqcup W \equiv C
\end{prooftree}
\qquad
\begin{prooftree}
\justifies
\vdash Ch \sqcup \ell \equiv Ch
\end{prooftree}
\end{gather*}
Per this, $\vdash I \leq W \leq C \leq Ch$.
These labels are associated with further foundational axioms:
\begin{gather*}
\begin{prooftree}
\vdash \Gamma \text{ valid}
\quad
\vdash \phi : \mathtt{Prop}
\justifies
\Gamma \vdash \neg\phi \vee \neg\neg\phi : W
\using\rulefont{\star}
\end{prooftree}
\qquad
\begin{prooftree}
\vdash \Gamma \text{ valid}
\quad
\vdash P : \alpha \rightarrow \beta \rightarrow \mathtt{Prop}
\justifies
\Gamma \vdash \fall{x}\xsts{y}P\ x\ y \longrightarrow \xsts{f}\fall{x}P\ x\ (f\ x) : Ch
\using\rulefont{\star\star}
\end{prooftree}
\end{gather*}
Here, the rule annotated with \rulefont{\star} induces Jankov's logic, or the logic of the Weak Excluded Middle \cite{JankovVA:calwealem}, a superintuitionistic logic weaker than classical logic, whilst the rule annotated \rulefont{\star\star} introduces the Axiom of Choice.
Each new axiom satisfies requisite implications between all other foundational axioms---recall that Choice implies every other axiom by Diaconescu's theorem, and acts as the upper bound of the taint-label ordering.

Care must be taken when adding new labels that the \emph{syntactic} lattice of taint-labels indeed remains a lattice and labels in this syntactic lattice must always reflect entailments in the underlying \emph{semantic} lattice relating different foundational axioms.
We call this reflection a \deffont{coherence} property.
Two taint-labels, $\{\ell, \ell'\} \subseteq \mathcal{L}$, are related by $\vdash \ell \leq \ell'$ whenever the axiom associated with label $\ell$ follows from the axiom associated with $\ell'$.
Specifically, here, Weak Excluded Middle is implied by Excluded Middle, hence $\vdash W \leq C$, but not \emph{vice versa}.
We will return to this subject later in \S\ref{sect.data.and.definitional.principles} when we discuss definitional principles.

Lastly, observe that the logic is ambivalent about \emph{which} of a collection of logically equivalent foundational axioms are added to the logic.
A large number of results that are equivalent to Choice are known, including Zorn's lemma, Tarski's theorem, and the fact that every surjection has an injective inverse.
Obviously, each of these results can be provided as a derived rule within the $Ch$ fragment of our logic.
However, each could also be introduced as separate foundational axioms---though perhaps awkwardly, for some---with their own dedicated taint-label.
Owing to the bi-implication between the two Choice-equivalents, and the fact that $\vdash \ell \leq \ell'$ and $\vdash \ell' \leq \ell$ implies $\vdash \ell \equiv \ell'$, this new label is derivably equal to the label $Ch$, and results in either world are therefore immediately available in the other.

\section{Data, and definitional principles}
\label{sect.data.and.definitional.principles}

For presentation purposes, we simply assert the existence of \deffont{strictly-positive data}, and \deffont{inductively-defined relations}, along with associated reasoning principles within our system, rather than constructing them via conservative extension.
The datatype $\mathtt{Bool}$ will prove useful, introduced by \deffont{constructors}, and with a \deffont{primitive recursor}, captured by:
\begin{displaymath}
\begin{tabular}{ccl}
$\mathtt{true}$, $\mathtt{false}$  &  \text{ with type } & $\mathtt{Bool}$ \\
$\mathtt{ite}$   &  & $\mathtt{Bool} \rightarrow \alpha \rightarrow \alpha \rightarrow \alpha$
\end{tabular}
\end{displaymath}
We again take pity on the reader, write $\mathtt{if}\ r\ \mathtt{then}\ t\ \mathtt{else}\ f\ \mathtt{end}$ instead of $((\mathtt{ite}\ r)t)f$, and continue to suppress type-substitutions to make terms well-typed.
We also assume axioms asserting that the constructors for $\mathtt{Bool}$ are \emph{free}, with all constructors distinct and injective, as well as axioms describing the equational properties of the recursor.
Finally, we also assume a \deffont{structural induction rule} to derive properties over elements of type $\mathtt{Bool}$.
All axioms live in $I$, the purely-constructivist world, and can be freely lifted into any other.

Sets are a useful data structure and endemic throughout modern mathematics.
Our logic provides us with a more fine-grained choice in how we define sets compared to HOL as we can define sets as predicates into either $\mathtt{Prop}$ or $\mathtt{Bool}$.
Both will prove useful---depending on the world within which we work---as the former is more useful for specifications and the latter for definitions, at least in all but the most classical of worlds.
We retain the \deffont{set} terminology for sets constructed from predicates over $\mathtt{Prop}$ and use \deffont{collection} for predicates over $\mathtt{Bool}$.

To introduce sets, we fix an additional type-former $\mathtt{Set}$ of kind $\star \Rightarrow \star$, so that $\mathtt{Set}\ \tau$ is a type for all types $\tau$, and set this as a type-synonym for $\tau \rightarrow \mathtt{Prop}$.
Assume an additional polymorphic constant, $-{\in}-$ with type $\alpha \rightarrow \mathtt{Set}\ \alpha \rightarrow \mathtt{Prop}$, write $x \in S$ and $\{ x:\tau \mid \phi \}$ as abbreviations for $S\ x$ and $\lam{x{:}\tau}\phi$, respectively.
Using this we can define further set operations with the expected types in the ``obvious'' way, for example:
\begin{displaymath}
\begin{tabular}{cclcl}
$\emptyset$ & & $\mathtt{Set}\ \alpha$ & & $\{ x : \alpha \mid \bot \}$ \\
$\mathtt{UNIV}$ & & $\mathtt{Set}\ \alpha$ & & $\{ x : \alpha \mid \top \}$ \\
$- \cup -$ &  \text{with type} & $\mathtt{Set}\ \alpha \rightarrow \mathtt{Set}\ \alpha \rightarrow \mathtt{Set}\ \alpha$ & \text{defined by} & $\lam{S}\lam{T}\{ x : \alpha \mid x \in S \vee x \in T \}$ \\
$\mathtt{cmpl}$ &  & $\mathtt{Set}\ \alpha \rightarrow \mathtt{Set}\ \alpha$ & & $\lam{S}\{ x : \alpha \mid \neg(x \in S) \}$
\end{tabular}
\end{displaymath}
Obviously, depending on the world, we obtain very different set theories from these definitions:
\begin{gather*}
\begin{prooftree}
\vdash \Gamma \text{ valid}
\quad
\vdash S, T : \mathtt{Set}\ \tau
\justifies
\Gamma \vdash S \cup T = T \cup S : I
\end{prooftree}
\qquad
\begin{prooftree}
\vdash \Gamma \text{ valid}
\justifies
\Gamma \vdash \mathtt{cmpl}\ \emptyset = \mathtt{UNIV} : I
\end{prooftree}
\qquad
\begin{prooftree}
\vdash \Gamma \text{ valid}
\justifies
\Gamma \vdash \mathtt{cmpl} \circ \mathtt{cmpl} = \mathtt{id} : C
\using\rulefont{\dagger}
\end{prooftree}
\end{gather*}

Here in the derived rule annotated with \rulefont{\dagger} we make implicit use of a \deffont{function composition} constant $- \circ -$ at type ($\alpha \rightarrow \beta) \rightarrow (\beta \rightarrow \gamma) \rightarrow \alpha \rightarrow \gamma$ and the \deffont{identity} function $\mathtt{id}$ at type $\alpha \rightarrow \alpha$, with standard definitions.
Note that this derived rule does not hold in the constructivist world, $I$, and is phrased in a way most convenient to make use of our logic's natural extensionality, as captured by the following derived rule:\footnote{We could further introduce a distinction between intensional and extensional reasoning by adjusting the $\eta$-axiom to introduce a new label for constructive, extensional reasoning, reinterpreting $I$, if so desired.}
\begin{gather*}
\begin{prooftree}
\Gamma \vdash f\ x = g\ x : \ell
\quad
x \notin fv(\Gamma)
\justifies
\Gamma \vdash f = g : \ell
\end{prooftree}
\end{gather*}
This extensionality flows through into our internal set theories, wherein two sets are considered equal if they contain the same elements, in all worlds, as is standard in pen-and-paper mathematics and other HOL variants:
\begin{gather*}
\begin{prooftree}
\Gamma \vdash x{:}\tau \in S \longleftrightarrow x{:}\tau \in T : \ell
\quad
x{:}\tau \notin fv(\Gamma)
\justifies
\Gamma \vdash S = T : \ell
\end{prooftree}
\end{gather*}
Returning to the distinction between $\mathtt{Prop}$ and $\mathtt{Bool}$, we have an embedding function $\mathtt{lift}$ of type $\mathtt{Bool} \rightarrow \mathtt{Prop}$, definable using the $\mathtt{ite}$ constant, and described piecewise by:
\begin{gather*}
\begin{prooftree}
\vdash \Gamma \text{ valid}
\justifies
\Gamma \vdash \mathtt{lift}\ \mathtt{true} = \top : I
\end{prooftree}
\qquad
\begin{prooftree}
\vdash \Gamma \text{ valid}
\justifies
\Gamma \vdash \mathtt{lift}\ \mathtt{false} = \bot : I
\end{prooftree}
\end{gather*}
Outwith the more classical worlds we are unable to obtain an inverse function of type $\mathtt{Prop} \rightarrow \mathtt{Bool}$, as we are prohibited from working by cases on elements of $\mathtt{Prop}$.
That, however, changes in the $Ch$ world, given access to the Axiom of Choice.
Introducing the relation $R : \mathtt{Prop} \rightarrow \mathtt{Bool} \rightarrow \mathtt{Prop}$ so that $R\ \top\ \mathtt{true}$ and $R\ \bot\ \mathtt{false}$ both hold, we have:
\begin{gather*}
\begin{prooftree}
\[
\[
\vdash \Gamma \text{ valid}
\leadsto
\justifies
\Gamma \vdash \fall{x}\xsts{y}R\ x\ y : C
\]
\justifies
\Gamma \vdash \fall{x}\xsts{y}R\ x\ y : Ch
\]
\quad
\[
\vdash \Gamma \text{ valid}
\justifies
\Gamma \vdash \fall{x}\xsts{y}R\ x\ y \longrightarrow \xsts{f}\fall{x}R\ x\ (f\ x) : Ch
\]
\justifies
\Gamma \vdash \xsts{f}\fall{x}R\ x\ (f\ x) : Ch
\end{prooftree}
\end{gather*}
The missing proof is established as a corollary of Excluded Middle, in $C$.
From this we obtain $\mathtt{drop} : \mathtt{Prop} \rightarrow \mathtt{Bool}$ with $\mathtt{lift}$ and $\mathtt{drop}$ provably mutually inverse.
The distinction between $\mathtt{Prop}$ and $\mathtt{Bool}$ thus collapses, as does the distinction between sets and collections.

Switching tack, one of Gordon's innovations in HOL over Church's Simple Theory of Types was the introduction of a definitional rule for ``carving out'' new types from old, using non-empty subsets of elements of a pre-existing, host type.
The introduction of this definitional principle in the HOL family of theorem proving systems was an advance over the earlier LCF family, which largely relied on axiomatisation to introduce new types.
Indeed, this mechanism is one way through which arbitrary datatypes---which we simply assumed by fiat above---can be constructed from scratch, as is actually done in many HOL implementations.

Interestingly, we can include a similar mechanism for introducing similar \deffont{subset types}, albeit it seems each world essentially has its own version, corresponding to the different conception of \emph{existence} present in each world.
For example, in the classical world $C$ we have the conception of existence as coinciding with the impossibility of non-existence:
\begin{gather*}
\begin{prooftree}
\Gamma \cup \{ \fall{x}\neg(\phi\ x) \} \vdash \bot : \ell
\justifies
\Gamma \vdash \xsts{x}\phi\ x : C
\end{prooftree}
\end{gather*}
This of course differs from the constructive notion of existence which requires an explicitly-constructed witness, as used within the constructivist world, $I$.
Given this, consider inductively defining \deffont{finiteness} for sets as a relation, $\mathtt{finite}$ of type $\mathtt{Set}\ \alpha \rightarrow \mathtt{Prop}$, by:
\begin{gather*}
\begin{prooftree}
\vdash \Gamma \text{ valid}
\justifies
\Gamma \vdash \mathtt{finite}\ \emptyset : I
\end{prooftree}
\qquad
\begin{prooftree}
\Gamma \vdash \mathtt{finite}\ S : \ell
\quad
\vdash S : \mathtt{Set}\ \alpha
\justifies
\Gamma \vdash \mathtt{finite}\ (S \cup \{ x{:}\alpha \}) : \ell
\end{prooftree}
\end{gather*}
Then we may ``carve out'' an explicit type of finite sets, $\mathtt{Fset}\ \alpha$, from the $\mathtt{Set}\ \alpha$ type using $\mathtt{finite}$.
To do this, we establish the existence of a set $S$ satisfying $\mathtt{finite}$, which can be done constructively, in $I$, with an explicit witness---say $\emptyset$---or classically, in $C$ or $Ch$.
Given this, we obtain \deffont{injection} and \deffont{projection} functions, $\mathtt{inj}$ and $\mathtt{proj}$, into and out of the new type, respectively, and mediated by the following pair of laws:
\begin{gather*}
\begin{prooftree}
\vdash \Gamma \text{ valid}
\justifies
\Gamma \vdash \mathtt{proj} \circ \mathtt{inj} = \mathtt{id} : \ell
\end{prooftree}
\qquad
\begin{prooftree}
\vdash \Gamma \text{ valid}
\quad
\vdash S : \mathtt{Set}\ \alpha
\justifies
\Gamma \vdash \mathtt{finite}\ S \longrightarrow \mathtt{proj}\ (\mathtt{inj}\ S) = S : \ell
\end{prooftree}
\end{gather*}
With these, we may ``lift'' functions defined on sets into the new type, and establish properties of these lifted functions by ``dropping down'' to the underlying host type used for the carve-out.
Here, we conjecture that the label $\ell$ appearing in the conclusions of the rules above should match that used to establish the existence of a witness, as part of the carve-out process.

Lastly, we speculatively consider the possibility of a \emph{new} definitional principle for taint.
Until now we have worked with an arbitrary bounded lattice which was specialised in the previous section.
The particulars of this specialisation are, as discussed, a matter of taste.
Instead, we could add a new principle for \emph{dynamically} adding taint, using proof \emph{within} the system itself to relate new taint to old, introducing an ``open world'' of foundational axioms.

Specifically, one rather prosaic interpretation of our taint is as a labour saving device: taints remove the need to manually thread axioms throughout proofs, removing an impediment to ergonomic working.
As discussed previously, from this the lattice structure of taint-labels emerges, reflecting entailments between the axioms represented by labels---our coherence property.
We therefore have the following \textbf{unwinding property} which allows us to ``drop'' a classical result back into the constructivist world, by relativising the (generalised) statement:
\begin{lemma}
\label{lemm.unwinding}
If $\Gamma \vdash \psi : C$ then $\Gamma \vdash (\fall{\phi}\phi \vee \neg\phi) \longrightarrow \psi : I$.
\end{lemma}
We take this as the notion of correctness for adding labels: in adding a new label, we remain coherent, with the taint-label lattice structure preserved, and foundational axioms associated with each taint-label \deffont{placed} with respect to axioms associated with every other label, by finding an existing label that is the least upper bound of the new label and each other label.\footnote{Here, we take $\top$ as the foundational axiom associated with the constructivist taint-label, $I$.}

However, we must take care to avoid becoming beguiled by an overly simple, purely implicational treatment of unwinding, as captured by Lemma~\ref{lemm.unwinding}, above.
Whilst this would work if axioms mentioned only ground types, as in the Excluded Middle, this falls apart when dealing with axioms that contain polymorphic types---the Axiom of Choice, for example.
Here, HOL's type system forces a distinction between formulae appearing in the statement of a theorem containing polymorphic types and the same formula appearing as the conclusion of an inference rule.
Types appearing within a formula become fixed, unable to vary across a proof, whilst types appearing in inference rules can vary arbitrarily, using the \nruletysubst{} rule.

As a result, we further conjecture one may place new taint labels in relation to existing ones in the lattice by temporarily ``turning off'' the taint system during the placing process.
For example, assuming our system thus-far only includes the taint-labels $I$ and $C$, we may add the new taint-label $W$, associated with Jankov's logic, by proving $\Gamma \vdash \top : I$ given a proof of $\Gamma \vdash \neg\phi \vee \neg\neg\phi : I$ and similarly proving $\Gamma \vdash \neg\phi \vee \neg\neg\phi : I$ given a proof of $\Gamma \vdash \neg\phi \vee \phi : I$.
These proofs must be carried out in the purely constructive fragment of the logic---without appeal to any other foundational axioms---and with all axioms temporarily inhabiting the ``constructivist world'', $I$.
From this, $\vdash I \sqcup W = W$ and $\vdash W \sqcup C = C$ and we obtain $\vdash I \leq W \leq C$, as expected.
This process can be easily generalised to consider multiple axioms associated with a taint-label, rather than a single one.

This proof-based approach merely establishes that a label acts as the upper bound of two other labels, rather than the \emph{least} upper bound.
To establish this, we also need to consider disproof, checking that axioms associated with any other label do not also act as an upper bound.
However, establishing this may require more elaborate means---for example, model-theoretic techniques---than can be captured from \emph{within} the logic.
If an ``open world'' of taint were to be accommodated, then any use of the definitional principle for adding new taint must be part of the logic's trusted base, with each usage audited to ensure that upper bounds used during the placing process are indeed least upper bounds.
We leave making this new definitional principle fully formal for further work.

\section{Pragmatics of encoding mathematics}
\label{sect.mathematics.and.smooth.infinitesimal.analysis}

Previously, we commented on the exclusion of any analogue of the Hilbert-style description operator, $\epsilon$.
As a result, we cannot \emph{define} concepts that would otherwise require $\epsilon$, for example a function's inverse, or the minimal element of an ordered set, as one can in HOL.
Instead, we must use \emph{relativisation}, wherein we first define a predicate $\mathtt{isinv}$ of type $(\alpha \rightarrow \beta) \rightarrow (\beta \rightarrow \alpha) \rightarrow \mathtt{Prop}$ with $\mathtt{isinv}\ f\ g$ asserting that the function $g$ is the inverse of $f$.
Definitions and theorem statements---otherwise phrased in terms of an inverse of a function, $f$---are parameterised by a function, $g$, with the explicit assumption $\mathtt{isinv}\ f\ g$ dischargeable by proof.
This is eminently less convenient, but admirably equitably progressive---everybody is disadvantaged equally, irrespective of belief or philosopho-religious stance.
Exploring ways to ameliorate this inconvenience via automation, or similar, is left for future work.

Further, one interesting aspect of our system is that it provides a strong incentive to try and phrase results in the weakest possible system that will suffice, as doing so maximises reuse of results between different worlds.
This is because everybody reasons over the same set of definitions, and also over the same types: there is no type of natural numbers, for example, specific to the constructivists, and distinct from those used by classical reasoners---\nrulelift{} makes permitted movements of results between worlds easy.
As a result, denizens of less classical worlds may need to introduce definitions, types, and new theorem statements in order to distinguish concepts that the more classically-inclined cannot observe, and which collapse as one assumes more-and-more classical styles of reasoning.
One example of this phenomenon is a concept already considered, namely finiteness of a set, which constructively is separable into several different related concepts.
The same observation also holds true for infinite sets, the definition of a field, various results in analysis, and so on.

Note that our choice of embedding the strictly-positive datatypes into our logic directly, in \S\ref{sect.data.and.definitional.principles}, is all that commits us to the existence of infinite sets, for example the set of natural numbers.
Most HOL implementations explicitly include an axiom asserting the existence of an infinite set, used to ``carve out'' other datatypes via a process of conservative extension.

Lastly, the particular brand of constructivism that we can reasonably capture is limited by the fact that, like HOL, our logic is \emph{impredicative}.
This makes our brand of constructivism similar in spirit to that implemented in the impredicative universes of Coq and Matita, but distinct from the \emph{predicative} constructivism implemented by Agda, or Coq and Matita's predicative universes.
The internal set theories of our logic also have a distinct flavour: powersets can be freely formed, putting us at odds with some other brands of constructive set theory where powerset formation is severely restricted~\cite{sep-set-theory-constructive}.
Nevertheless, we consider our constructivism as similar in spirit to the informal constructivism of mainstream mathematics.
Here, ``constructive'' is taken to mean working without Excluded Middle, Choice, or similar, and the description of explicit witnesses in existence proofs.

\section{Implementation}
\label{sect.computer.implementation}

We have written a prototype LCF-style~\cite{milnerR:usemactairp} proof-checking kernel implementing our logic\footnote{Our code is open-source and available at \url{https://github.com/martinberger/hol-c}.}, implemented in \emph{Scala 3}.
As is typical, derivation trees are captured by an abstract type---a \emph{class} of $\mathtt{Thm}$ objects in our case.
We also support basic backwards-directed proof, driven by \emph{tactics}, with each inference rule and axiom of our Natural Deduction, and certain useful derived rules, associated with a basic tactic.
Our prototype system also provides a system of \emph{tacticals}---or higher-order tactic-valued functionals---for building larger, more complex tactics from simpler building blocks.
In our system are implemented as a \emph{deep embedding} of a tactic DSL~\cite{SchmidtDA:pronotftr,MartinAP:taccalcav,MartinA:macasstpfse} interpreted by the system to obtain an action on proof-states.

With this, we can for example demonstrate that our system is able to correctly certify that \emph{Peirce's Law} resides in the ``classical world'' by proving the conjecture $\{\} \vdash (\phi \rightarrow \psi) \rightarrow \phi \rightarrow \phi : C$.
This follows from the classical \emph{reductio ad absurdum} principle, as captured by the \nruleraa{} derived rule, with our system supplying an explicit tactic associated with this principle.
Despite the overall result being classical, subproofs only require purely-constructive reasoning.
Consequently, these purely-constructive subproofs can be explicitly lifted into the ``classical world'', $C$, using a tactic that reverses the action of the \nrulelift{} rule, with subgoals dropped down into a user-specified world lower in the taint-lattice, per the derived ordering, than the current subgoal, and failing if the user-supplied world fails to satisfy this condition.

    

\section{Conclusions}
\label{sect.conclusions}

Mathematicians often perform temporary \emph{context switches} between different foundational systems when checking a proof, with explicit invitation to perform these context switches marked by informal narrative notes accompanying a proof or other mathematical text.
Yet, extant proof-checking software provides relatively poor support for switching between different modes of reasoning, making the formalisation of certain mathematical material cumbrous.

In some HOL implementations---HOL4, for example---theorem objects are tagged in order to track axiom usage---see the use of the $\mathtt{Tag.merge}$ function in the implementation of the HOL4 kernel, for example.
This taint is not exposed as part of the logic, and is merely an implementation detail.
Our innovation is essentially ``running with'' this idea by working taint into the fabric of the logic itself, with the effect of ``carving up'' the logic into distinct sublogics, sat side-by-side, of differing expressivity.

Whilst in some sense our taint-system is a labour saving device, obviating the need to thread axioms throughout proofs, interestingly it also looks and acts a lot like a static analysis or bare-bones typing system for HOL proofs.
Indeed, the movement of results between subsystems of our logic is mediated by a new inference rule strongly reminiscent of a subtyping or subsumption rule~\cite[Section 15.1 ``Subsumption'']{pierce-types-2002}.
Na\"ively, our taint also looks somewhat reminiscent of a form of \emph{modality}~\cite{DBLP:books/el/07/BBW2007}, though we caution the reader against this interpretation.
The formula language of our logic---as introduced in \S\ref{sect.a.more.pluralistic.hol}---is identical to that of standard HOL.
As a result, $\phi : \ell$ is \emph{not} a formula within our system for any $\ell \in \mathcal{L}$.

Our system is similar to a Labelled Deduction system for HOL, \emph{à la} Gabbay~\cite{gabbay-labelled}.
We are not aware of any similar labelling for HOL, though there exists a wide body of work applying labelling to relatively quotidian logics, like classical and intuitionistic first-order logics~\cite{DBLP:journals/sLogica/GabbayR97}, and the more exotic, like substructural~\cite{DBLP:journals/igpl/BrodaFR99, DBLP:journals/jar/DAgostinoG94}, modal~\cite{DBLP:journals/logcom/MarinMS21, DBLP:journals/logcom/BasinMV97}, and temporal logics~\cite{DBLP:journals/sLogica/Indrzejczak03}.
This resource can be mined for inspiration for future work: focussing on substructural logics, specifically, labelling has been previously applied to produce similar parametric systems for an array of substructural logics, with labels taken from a particular algebra which can be varied---like our lattice---to obtain different substructural logics.\footnote{In somewhat related work, Quantitative Type Theory~\cite{AtkeyR:synsemoqtt} and its predecessor systems~\cite{McBride2016} also uses an algebra---specifically a \emph{resource semiring} to smoothly integrate linear logic and dependent type theory.}

We speculate that our taint idea can also be used to further ``carve up'' HOL in a similar fashion to consider substructural fragments of HOL.
Here we focussed on adding or removing \emph{axioms}, substructural logics, however, restrict the use of different \emph{structural rules} within the Natural Deduction relation.
In our system these rules are largely tacit, barring \nruleweak, owing to our use of sets for contexts.
They could however be surfaced by modelling contexts as lists of assumptions, instead, and introducing explicit rules to manage and manipulate assumptions.
Structural rules can then also introduce taint, for example:
\begin{gather*}
\begin{prooftree}
\Gamma \vdash \psi : \ell
\quad
\vdash \phi : \mathtt{Prop}
\justifies
\phi {::} \Gamma \vdash \psi : \ell \sqcup Wk
\end{prooftree}
\end{gather*}
(Here $- {::} -$ is the \emph{list cons} operator, which appends an assumption onto the front of the context $\Gamma$, and $Wk$ is a dedicated taint-label associated with the weakening structural rule.)

This is a potentially more far-reaching change than we have introduced in the main body of our paper.
Restricting the use of structural rules can cause the standard logical constants and connectives to devolve into different subfamilies of connectives---linear logic, for example, decomposes the standard connectives into multiplicative and additive families.

\newpage

\bibliography{bib} 

\end{document}